\documentclass[aps,pra,twocolumn,superscriptaddress,preprintnumbers,longbibliography]{revtex4-2}

\usepackage[colorlinks=true, allcolors=blue]{hyperref}
\usepackage{upgreek} 
\usepackage{natbib}
\usepackage{graphicx}
\usepackage{comment}
\usepackage{amsmath}
\usepackage{amssymb}
\usepackage{ulem}

\begin{document}

\title{Cabibbo-Kobayashi-Maskawa unitarity deficit reduction via finite nuclear size}

\author{Mikhail Gorchtein}\email{gorshtey@uni-mainz.de}
\affiliation{
Institut f\"ur Kernphysik, Johannes Gutenberg-Universit\"at Mainz, 55128 Mainz, Germany
}
\affiliation{
PRISMA$^+$ Cluster of Excellence, Johannes Gutenberg-Universit\"at Mainz, 55128 Mainz, Germany
}

\author{Vaibhav Katyal}
\affiliation{
Atomic, Molecular and Optical Physics Division, Physical Research Laboratory, Navrangpura, Ahmedabad 380009, India and \\  
Indian Institute of Technology Gandhinagar, Palaj, Gandhinagar 382355, India
}

\author{B. Ohayon}\email{bohayon@technion.ac.il}
\affiliation{
The Helen Diller Quantum Center, Department of Physics,
Technion-Israel Institute of Technology, Haifa, 3200003, Israel
}

\author{B. K. Sahoo}\email{bijaya@prl.res.in}
 \affiliation{
Atomic, Molecular and Optical Physics Division, Physical Research Laboratory, Navrangpura, Ahmedabad 380058, Gujarat, India
}

\author{Chien-Yeah Seng}\email{seng@frib.msu.edu}
\affiliation{Facility for Rare Isotope Beams, Michigan State University, East Lansing, MI 48824, USA}
\affiliation{Department of Physics, University of Washington,
	Seattle, WA 98195-1560, USA}

\date{\today}

\preprint{NT@UW-25-3}

\begin{abstract}
We revisit the extraction of the $|V_{ud}|$ CKM matrix element from the superallowed transition decay rate of $^{26m}$Al$\rightarrow$$^{26}$Mg, focusing on finite nuclear size effects. 
The decay rate dependence on the $^{26m}$Al charge radius is found to be four times higher than previously believed, necessitating precise determination.
However, for a short-lived isotope of an odd $Z$ element such as $^{26m}$Al, radius extraction relies on challenging many-body atomic calculations. We performed the needed calculations, finding an excellent agreement with previous ones, which used a different methodology. This sets a new standard for the reliability of isotope shift factor calculations in many-electron systems.
The $\mathcal{F}t$ value obtained from our analysis is lower by $2.2\,\sigma$ than the corresponding value in the previous critical survey, resulting in an increase in $|V_{ud}|^2$ by $0.9\,\sigma$. 
Adopting $|V_{ud}|$ from this decay alone reduces the CKM unitarity deficit by one standard deviation, irrespective of the choice of $|V_{us}|$.

\end{abstract}

\maketitle

The Cabibbo–Kobayashi–Maskawa (CKM) matrix is a cornerstone of the electroweak sector of the Standard Model (SM)~\cite{Cabibbo:1963yz,Kobayashi:1973fv}. It contains nine elements $V_{ij}$ corresponding to pairs of different quark flavors $(u,d,c,s,t,b)$.
Weak universality implies that the CKM matrix is unitary, resulting in the relation $\sum_j |V_{ij}|^2=1$.
Conversely, a statistically significant departure from unitarity implies the breaking of weak universality, constituting new physics beyond the SM.

The squared sum of the elements in the first row, $|V_{ud}|^2+|V_{us}|^2+|V_{ub}|^2$, is measured to a precision of $10^{-4}$ and is found to fall short of unity by $2-3\,\sigma$~\cite{2020-HT, 2024-LQCD, 2023-CKM,2023-CKMrev}, hinting at either the existence of new physics or an underestimated uncertainty.
This CKM unitarity deficit is one of several persistent anomalies in the electroweak sector, which are garnering attention by the community (see e.g.~\cite{2024-Anom}).

As $|V_{ub}|^2\approx10^{-5}$ is too small to be currently relevant and the status of $|V_{us}|$ (or $|V_{us}|/|V_{ud}|$) is driven mainly by incompatibilities between the experimental results~\cite{2023-CKM}, we focus on $|V_{ud}|$ in this work.
The most accurate determination of it is based on the weighted average of the transition rates ($ft$ values) of superallowed beta decays.
In recent years, theory has driven the uncertainty eventually \textit{increasing} it.
Of all measured superallowed decays, the $^{26m}$Al$\rightarrow^{26}$Mg transition has the smallest uncertainty and thus dominates the determination of $|V_{ud}|$.
Due to the high accuracy of the $ft$ value below 0.1\%, various small corrections need to be carefully estimated.

In many tests of the Standard Model the precision is only warranted if certain hadronic properties are known with the precision that matches or exceeds the experimental one. Oftentimes, our understanding of the theory of the strong interaction, quantum chromodynamics (QCD), is not yet sufficient to reach such precision due to its non-perturbative nature at low energies. Data-driven approaches are designed to overcome this limitation by taking the needed hadronic properties from other data if this guarantees higher accuracy. In doing so, one has to keep in mind that quantities derived from data rely on a set of theoretical assumptions that must match those used in the precision test of the SM at hand. The case of the magnetic moment of the muon nicely illustrates this issue. 
The 2020 data-driven prediction of the SM theory~\cite{Aoyama:2020ynm} has been challenged by the lattice QCD calculation~\cite{Borsanyi:2020mff} and by the new data~\cite{CMD-3:2023alj}, raising concerns about possible flaws in the analysis of data or their use in the calculations. As a result, the new value of the SM theory shifted~\cite{Aliberti:2025beg} and is now in agreement with the final experimental result~\cite{Muong2025}. 

In case of nuclear $\beta$ decays, it is the nuclear charge radii that play a major role in predicting the expected decay rate and must be established to the order of 0.1\%.
Nuclear-theory calculations currently only provide a few percent accuracy, and it is common to use charge radii deduced from measurements. To deduce them from, e.g., measured frequencies of atomic transitions, theoretical ingredients are unavoidable and must be controlled. 

In this Letter, we revisit the extraction of $|V_{ud}|$ from the $^{26m}$Al$\rightarrow^{26}$Mg decay using the data-driven strategy outlined in Refs.~\cite{Seng:2022epj,Seng:2023cgl}. A recent laser spectroscopy measurement of the $^{26m}$Al$\,-\,^{27}$Al isotope shift at the ISOLDE and IGISOL facilities~\cite{2023-Al}
reported the value of the charge radius of the $^{26m}$Al isomer, $r(^{26m}\text{Al})=3.130(15)\,$fm. This value is at variance with the value $r(^{26m}\text{Al})=3.040(20)\,$fm assumed previously~\cite{2020-HT}. 

\begin{figure}[tbp]
    \centering
    \includegraphics[width=0.95\linewidth]{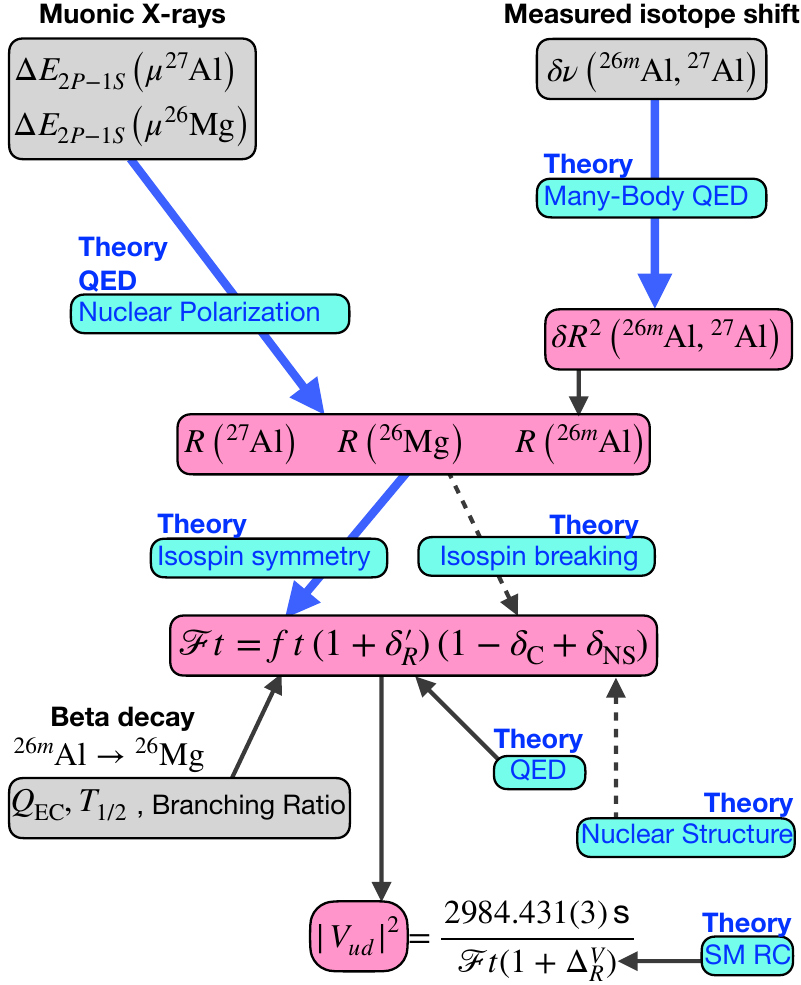}
    \caption{Schematic representation of the approach taken in this work to connect the measurements of the nuclear $^{26m}$Al$\,\rightarrow^{26}$Mg beta-transition to the atomic transitions used to determine the nuclear charge radii and to $V_{ud}$. A detailed explanation of all ingredients is given in the main text.
    }
    \label{fig:Scheme}
\end{figure}

To study the impact of this measurement on the $V_{ud}$ extraction from the $^{26m}$Al$\,\rightarrow^{26}$Mg superallowed transition, we follow the path schematically shown in Fig.~\ref{fig:Scheme}. 
The gray boxes represent purely experimental input: the aforementioned isotope shift measurement, the X-ray measurements of atomic transitions in muonic $^{27}$Al and $^{26}$Mg, and finally the $^{26m}$Al$\,\rightarrow^{26}$Mg transition itself. The blue boxes display the theory ingredients necessary to arrive at the constants of interest, highlighted by pink boxes: nuclear radii, the $\mathcal{F}t$-value, and $V_{ud}$. 
The thin solid black arrows correspond to the connections that are firmly under control. The dashed black arrows refer to theoretical calculations that are not addressed here and are taken from Ref.~\cite{2020-HT}. 

Finally, the thick solid blue arrows show the connections that are explicitly assessed in this work. Each entails a computation in a separate field. Because performing all three tasks in a single publication would have been too involved, we presented these calculations in a series of sister publications~\cite{2024-mirror, 2025-MG}, where all necessary details are given, and combined these results in this Letter for a concise and comprehensive analysis.
In addition to its direct impact on the $^{26m}$Al$\rightarrow^{26}$Mg transition, our work also serves to lay out a clear set of theoretical criteria, as depicted in Fig.~\ref{fig:Scheme}, which must be satisfied by all superallowed nuclear decays from which the extracted
$V_{ud}$ may be considered reliable.

\textit{$|V_{ud}|$ from superallowed $0^+\rightarrow0^+$ nuclear beta decays} may be extracted from measurements of three quantities: the energy release $Q_\mathrm{EC}$, the total half-life $T_{1/2}$, and the branching ratio; the latter two combine to yield the partial half-life $t$~\cite{2020-HT},
\begin{equation}\label{eq:Vud}
    |V_{ud}|^2= \frac{2984.431(3)\,\text{s}}{\mathcal{F}t(1+\Delta_R^V)}.
\end{equation}
The numerator is derived from the measured muon lifetime~\cite{2013-Mulan} combined with the relevant fundamental constants, notably the muon-to-electron mass ratio~\cite{2019-Eides}. 
$\Delta_R^V$ is the universal (nucleus-independent) single-nucleon \textit{inner} electroweak radiative correction. It has recently been under intense scrutiny~\cite{Seng:2018yzq,Czarnecki:2019mwq,Seng:2020wjq,Shiells:2020fqp,Hayen:2020cxh,Gorchtein:2023srs, 2024-LQCD}, and here we take the value from a recent review~\cite{Gorchtein:2023srs}, $\Delta_R^V=2.479(21)\%$.

$\mathcal{F}\equiv f(1+\delta_r')(1+\delta_\text{NS}-\delta_\text{C})$ is a product of three non-universal (nucleus specific) corrections $\delta_r'$, $\delta_\text{NS}$ and $\delta_\text{C}$, and the \textit{statistical rate function} $f$. 
$\delta_r'(A=26)=1.478(7)\%$~\cite{Towner:2007np} is the so-called \textit{outer} radiative correction for a point charge distribution.
It is calculable with quantum electrodynamics (QED)~\cite{Sirlin:1967zza,Sirlin:1986hpu,Sirlin:1986cc} and depends only on the charge $Z$ and the maximum energy, with uncertainty at the
$\mathcal{O}(Z^2\alpha^3)$ level. Following the prescription in Ref.~\cite{2020-HT}, we estimate its uncertainty conservatively to be one third of the entire $Z^2\alpha^3$ contribution. 
The nucleus-dependent \textit{inner} radiative correction $\delta_\text{NS}$ is not strongly dependent on the radius and is another quantity currently under scrutiny. In the absence of \textit{ab initio} calculations of $\delta_\text{NS}$ for $A=26$ we adopt the value $-0.019{(19)(47)}\%$ from the nuclear shell model~\cite{2020-HT}, with the single-nucleon contribution corrected for binding effects in a simple free Fermi gas picture~\cite{Seng:2018qru}. 
For light nuclear transitions ($A=10,14$), this approach is consistent with more sophisticated \textit{ab initio} calculations~\cite{Gennari:2024sbn,Cirigliano:2024msg}.
The first uncertainty in $\delta_\text{NS}$ is statistical and appears in the individual $\mathcal{F}t$ values given in Ref.~\cite{2020-HT}. The second uncertainty is systematic, added to the average $\overline{\mathcal{F}t}$ value given in Ref.~\cite{2020-HT}.

The departure of the Fermi matrix element from its isospin limit is encoded in $\delta_\text{C}$. It originates predominantly from Coulomb repulsion between protons in the nucleus. \textit{Ab initio} theory of $\delta_\text{C}$ is still in the preliminary stage~\cite{Caurier:2002hb,Martin:2021bud,Seng:2023cvt}, so for this work we use nuclear corrections from the traditional shell model approach of Towner and Hardy~\cite{Towner:2007np} (both indicated by the dashed arrows in Fig.~\ref{fig:Scheme}). In that formalism, the dependence of $\delta_\text{C}$ on nuclear charge radii is encoded in the parameters of the Woods-Saxon potential for the parent nucleus.

In their initial calculation~\cite{Towner:2002rg}, Towner and Hardy adopted $r(^{26m}\text{Al})=3.040(20)\,$fm, to arrive at $\delta_\text{C}=0.310(18)\%$~\cite{2020-HT}.
This radius was estimated by extrapolating from neighboring nuclei.
Recently, laser spectroscopy measurements were performed on the Al chain at the ISOLDE and IGISOL facilities~\cite{2023-Al}, returning $r(^{26m}\text{Al})=3.130(15)\,$fm that results in an update $\delta_\text{C}=0.340(17)\%$~\cite{2023-Al}. 
Based on these results, and since a small variation in $r(^{26m}$Al) tends to produce a linear relationship with $\delta_\text{C}$, we can parametrically express
\begin{equation}\label{eq:deltac}
    \delta_\text{C}\approx 0.310(17)\%+0.33\% \left[r(^{26m}\text{Al})/\text{fm}-3.040\right],
\end{equation}
with uncertainty of the first term independent of the radius.

If finite nuclear size (FNS) effects only affected $\delta_\text{C}$, the accuracy goal of $0.01\%$ in the $\mathcal{F}t$ value would directly translate to an accuracy goal of $0.030$~fm in $r(^{26m}\text{Al})$. 
However, we argue here that a significantly more stringent accuracy goal is required by the finite-size dependence of $f$.
The latter is commonly parameterized as~\cite{Hardy:2008gy}
\begin{equation}\label{eq:rate}
    f=\frac{1}{m^5}\int_{m}^{E_0}pE(E_0-E)^2F(E)C(E)Q(E)R(E)r(E)dE~,
\end{equation}
where $E=E_e$ and $p=|\vec{p}_e|$ are the positron energy and 3-momentum, respectively, and $E_0=M_f-M_i$ is the positron end-point energy. Of the functions that enter Eq.~(\ref{eq:rate}), only two depend on nuclear radii: The Fermi function $F(E)$ which accounts for the Coulomb interaction between the outgoing positron and the daughter nucleus~\cite{Fermi:1934hr}, and the shape factor $C(E)$~\cite{behrens1982electron} which accounts for the spatial distribution of the beta decay probability. 
The latter is encoded in a ``charged weak'' distribution $\rho_\text{cw}(r)$.
Using isospin symmetry, it can be related to the electric charge distributions of the members of the isotriplet~\cite{Seng:2022inj,Seng:2023cgl}, 
$\rho_\text{cw}(r)^{26}\approx 13\rho_\text{ch}^{^{26m}\text{Al}}(r)-12\rho_\text{ch}^{^{26}\text{Mg}}(r)$.

The second moment of the charge distribution, the RMS charge radius, is dominant.
We compute $f$ using the formalism in Ref.~\cite{Seng:2023cgl} considering small variations of $r(^{26m}\text{Al})$ and $r(^{26}\text{Mg})$ around $(3.040,3.0337)\,$fm, affecting both $F(E)$ and $C(E)$, and obtain the following approximate formula:
\begin{eqnarray}\label{eq:f}
    f&\approx 478.45(14)&-4.0\left[r(^{26m}\text{Al})/\text{fm}-3.0400\right]\nonumber\\
    &&+3.3\left[r(^{26~}\text{Mg})/\text{fm}-3.0337\right]~,\label{eq:fapprox}
\end{eqnarray}
where the uncertainty of the first term is a combination of those of the experimental $Q_\text{EC}$ value and, to a lesser extent, the atomic screening and the higher-order parameters of the nuclear charge distribution.
Note that Eq.~(\ref{eq:f}) reproduces $f_\text{new}$ given in Table~III of Ref.~\cite{Seng:2023cgl} when taking $r(^{26m}\text{Al})=3.130\,$fm and $r(^{26}\text{Mg})=3.0337\,$fm.

Comparing Eqs.~(\ref{eq:deltac}) and~(\ref{eq:f}) reveals that the dependence of $f$ on the radius, which is not taken into account in Ref.~\cite{2023-Al}, is three times stronger than that of the $\delta_\text{C}$.
The two effects add constructively so that in total
\begin{eqnarray}\label{eq:F}
    \mathcal{F}&\approx 483.92(8)(14)({9})&-5.6\left[r(^{26m}\text{Al})/\text{fm}-3.0400\right]\nonumber\\
    &&+3.3\left[r(^{26~}\text{Mg})/\text{fm}-3.0337\right],\nonumber\\
\end{eqnarray}
has a 4 times stronger dependence on the radius than $\delta_\text{C}$.
The uncertainties in the first term come (in order) from the first term in Eq.~(\ref{eq:deltac}), the first term in Eq.~(\ref{eq:fapprox}), and the statistical uncertainty in $\delta_\text{NS}$. Two remaining uncertainties, namely the one in $\delta_r'$ and the systematic uncertainty in $\delta_\text{NS}$, will be added at a later stage to facilitate a clear comparison with the values in the latest critical survey~\cite{2020-HT}.

Eq.~(\ref{eq:F}) shows that to limit the uncertainty contribution of the FNS to $|V_{ud}|$ to the level of $10^{-4}$ one needs to determine $r(^{26m}\text{Al})$ better than $0.28\%$, which is challenging for a rare isotope with an odd number of protons. 
The accuracy goal of $0.48\%$ for $r(^{26}\text{Mg})$ is much less stringent and is also easier to achieve, as it is a stable nucleus whose radius was measured directly with muonic atom x-ray spectroscopy~\cite{1992-MuX}.
Next, we revisit the extraction of both radii.
Using the updated radii, we revise the statistical rate function $f$ using the data-driven approach of Refs.~\cite{Seng:2022epj,Seng:2023cgl}. This allows us to obtain a revised $\mathcal{F}t$-value for this transition, from which $|V_{ud}|$ can be deduced.
With this, we can address the implications for the top-row CKM unitarity test.

\textit{Reference radii---}
The radius of a short lived nucleus is extracted from a combination of a \textit{reference radius}, typically measured with a combined analysis of muonic-atom x-ray spectroscopy and electron scattering~\cite{1995-Fricke}, and a radius difference extracted from optical isotope shifts (ISs, see e.g.,~\cite{2023-spec}).
We first focus on the reference radii. 
These have been recently reevaluated to be $r(^{26}\text{Mg})=3.0301(32)\,$fm and $r(^{27}\text{Al})=3.0601(34)\,$fm accounting for systematic uncertainties in the nuclear shapes~\cite{2024-mirror}.
However, Ref.~\cite{2024-mirror} relied on the nuclear polarization correction quoted in Fricke and Heilig's compilation~\cite{2004-intro}: Npol$(^{26}\text{Mg})=33(10)\,$eV and Npol$(^{27}\text{Al})=40(12)\,$eV. 
In this work, we update these values based on a recent evaluation by one of us~\cite{2025-MG}:
Npol$(^{26}\text{Mg})=36(3)\,$eV and Npol$(^{27}\text{Al})=48(5)\,$eV.
The main difference from the previous estimates could be traced back to the added hadronic effects (``nucleon polarization'').
The resulting reference radii are
$r(^{26}\text{Mg})=3.0307(25)\,$fm and $r(^{27}\text{Al})=3.0614(29)\,$fm. These radii extractions is one of the original results of this work, indicated in Fig.~\ref{fig:Scheme} by thick solid blue arrows. 

\textit{Differential radius and IS factors---}
Having determined the reference radii, we can obtain $r(^{26m}\text{Al})$ from the mean-squared difference $\delta(r^2)_{26m,27}\equiv r^2(^{27}\text{Al})-r^2(^{26m}\text{Al})$ using a measured IS $\delta\nu_{26m,27}$, and the linear-response equation
\begin{equation}\label{eq:IS}
    \delta\nu_{26m,27}\approx K\mu_{26m,27}+F\delta(r^2)_{26m,27}.
\end{equation}
The first term on the right is the mass shift (MS) proportional to $\mu_{26m,27} \equiv (M_{27} - M_{26m})/(M_{26m} M_{27})$, with nuclear mass, $M$ and the MS factor, $K$. 
$K$ may further be divided into a one-body part, $K^\text{NMS}$ and a two-body part, $K^\text{SMS}$, where the superscripts refer to `normal mass shift' and `specific mass shift', respectively.
The second term is the field shift (FS) proportional to the FS factor, $F$.
Refinements to Eq.~(\ref{eq:IS}), e.g. from mixed terms~\cite{2025-Rec} and polarizability~\cite{2021-NPOLe}, are completely negligible in the accuracy goals of this work (see below).

In a monoisotopic element such as Al, the IS factors $F$ and $K$ cannot be deduced from measured radii differences and therefore must be calculated with many-body atomic theory.
More than $90\,\%$ of $\delta\nu_{26m,27}$, measured with the $^2P_{3/2}$--$^2S_{1/2}$~\cite{2021-Al} transition, originates from MS.
Therefore, and considering that the measurement accuracy is $0.9\,\%$, the objective of the accuracy of calculating $F$ is 10\%.
This is well within the capabilities of several many-body methods~\cite{2024-review}.
On the other hand, $K$ should be calculated to a relative precision better than $1\,\%$. This is a highly challenging accuracy goal for a neutral many-electron system where electron correlations are prominent~\cite{2012-Co, 2024-review}.
Due to this difficulty, independent calculations of $K$ using different methods or with values obtained semi-empirically generally agree only to within a few percent (see e.g.~\cite{2012-Co, 2022-Cd, 2022-Na, 2023-optical, 2024-K, Leonid, 2023-Si, 2023-Tl, 2007-Kozlov}). 
Usually, these deviations are larger than the `internal' uncertainties assessed by the individual groups.

In atomic aluminum, a recent study used the relativistic coupled-cluster method (RCC) to determine IS factors with sub-percent accuracy~\cite{Leonid}.
It used the DIRAC software package~\cite{DIRAC24} that adopts the group symmetry approach in the Cartesian coordinate system to perform atomic calculations.
We refer to it as the `molecular code'.
Although this calculation rigorously treats QED effects and may be considered the state of the art, it must be validated using an independent method of comparable accuracy.
In view of this, we calculate IS factors using the RCC method as implemented in the spherical coordinate system and compare them with the values in the literature. We refer to this as the ``atomic code".
Detailed results of our calculation are given in the Appendix~\ref{appB} with more information on our methods found in recent publications~\cite{2024-K,2024-review}.
This calculation is another original contribution of this work and is indicated by the thick solid arrow in Fig.~\ref{fig:Scheme}.

The resulting total MS factor for the transition of interest is $K=-240.8(1.2)\,$GHz\,u. Its uncertainty is mainly ``external", which means that it originates from a comparison of the two different independent calculation methods.
Combining this value of $K$ with $F=77.8(5)\,$MHz\,fm$^2$, and the measured $\delta\nu_{26m,27}=377.5(3.4)~$MHz~\cite{2023-Al} returns 
$\delta(r^2)_{26m,27}=0.461(49)\,$fm$^2$, with uncertainty now predominantly experimental. The agreement between the two very different calculation methods lends credibility to the differential radius obtained.

Combining both absolute and differential radii returns $r(^{26m}\text{Al})=3.136(8)\,$fm.
Its uncertainty is slightly lower than that given in~\cite{2023-Al, Leonid} due to the different estimations of the charge distribution dependence~\cite{2024-mirror} and our nuclear polarization correction. The uncertainty is now dominated by the optical experiment~\cite{2023-Al}, motivating higher-precision measurements of the same transition or exploring transitions with better sensitivity to the FS. Another avenue for improvement is to extend muonic-atom measurements to a microgram target of $^{26}$Al~\cite{2023-Micro, 2023-Implanted}.

\textit{Updated $\mathcal{F}$t value and $|V_{ud}|$---}
Plugging in both updated radii in Eq.~(\ref{eq:F}) returns $\mathcal{F}=483.37(18)(5)$, where the first and second uncertainties are from the FNS-independent and FNS-dependent contributions, respectively.
The partial half-life, $t(^{26m}\text{Al})=6351.24(55)\,$ms can be found in Ref.~\cite{2020-HT}. 
The corresponding individual corrected rate is $\mathcal{F}t=3070.0(1.2)_\text{stat}\,$s (the error is ``statistical'').
 It can be compared with the value of $3072.4(1.1)_\text{stat}\,$s in Table XVI of Ref.~\cite{2020-HT} obtained before the new $r(^{26m}\text{Al})$ measurement, as well as $3071.4(1.0)_\text{stat}\,$s in Table II of Ref.~\cite{2023-Al} where only the impact of the new $r(^{26m}\text{Al})$ on $\delta_\text{C}$ was taken into account.
 Our recommended value differs from the result in Refs.~\cite{2020-HT} and \cite{2023-Al} by $2.2\,\sigma$ and $1.4\,\sigma$, respectively, indicating the large impact of FNS effects.

This work demonstrates strong dependence of $\mathcal{F}$ on the absolute value of the RMS charge radius of the parent nucleus that undergoes a superallowed beta decay. 
In most cases, this radius was not measured, or the uncertainty in its value, dominated by many-body atomic calculations, is prohibitively large.
In light of this, we opt not to average the individual $\mathcal{F}t$ values of different transitions, but for now rely only on the one that we scrutinized.
Adding back the uncertainty of $\delta_r'$ and the systematic uncertainty of $\delta_\text{NS}$, we recommended the final value of
$\mathcal{F}t=3070.01(1.23)_\text{stat}(0.20)_{\delta_r'}(1.44)_{\delta_\text{NS}}\,$s from the investigated transition after taking into account the entire uncertainty.
This may now be compared with the average value $\overline{\mathcal{F}t}=3071.96(0.56)_\text{stat}(0.36)_{\delta_r'}(1.73)_{\delta_\text{NS}}\,$s given in Ref.~\cite{2023-Al}. The two have comparable total uncertainties, but the former is about $1\,\sigma$ smaller in its central value.

\begin{figure}[tbp]
    \centering
    \includegraphics[trim=5.5cm 2mm 5.8cm 1cm,clip,width=0.99\linewidth]{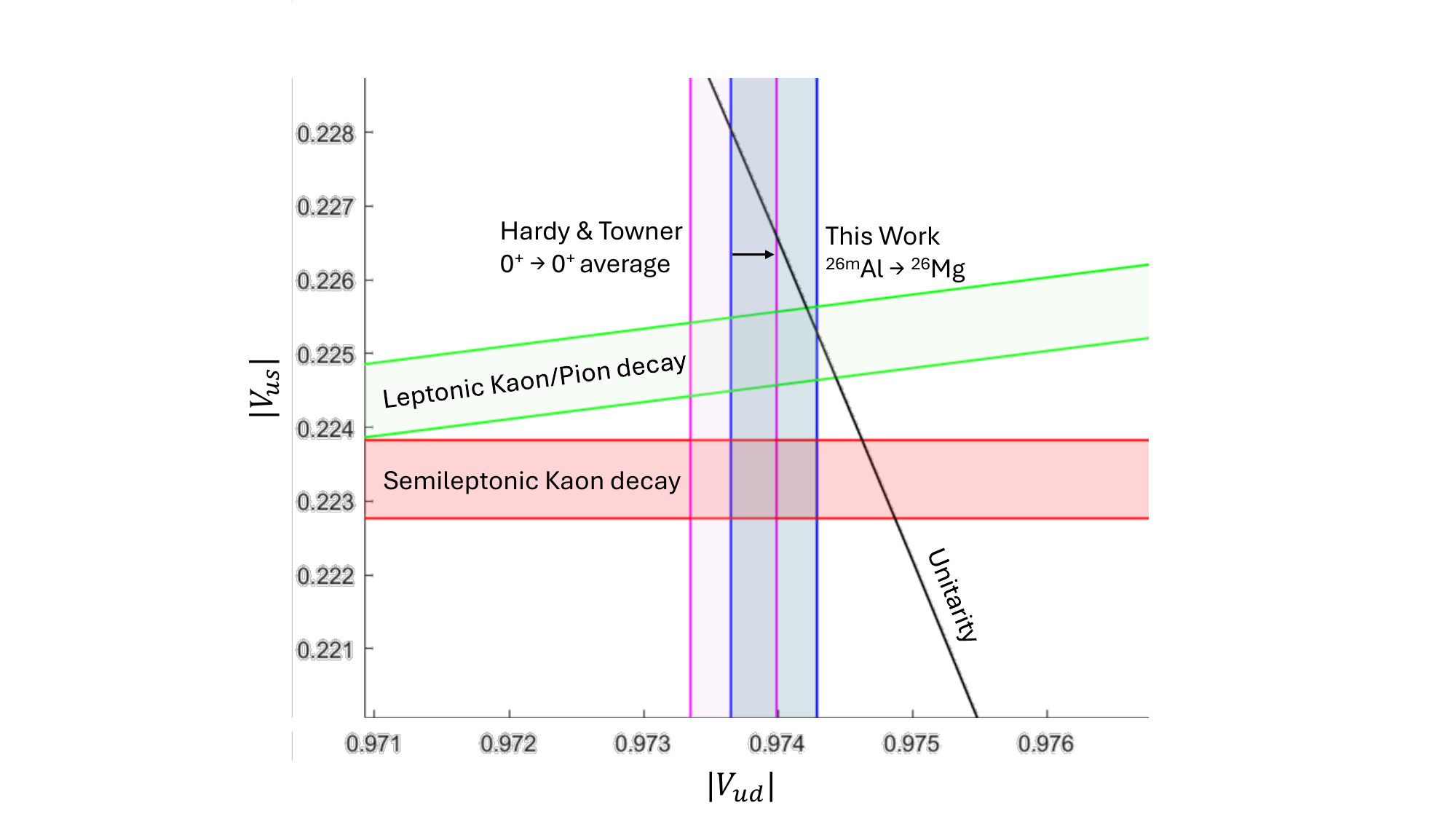}
    \caption{
Effect of the updated value of $|V_{ud}|$ form this work on the CKM unitarity test. The bands span the 67\% confidence interval.
    }
    \label{fig:CKM}
\end{figure}

The updated value $|V_{ud}|^2=0.94861(62)$ solely from this system has a similar uncertainty and is $0.9\,\sigma$ above $|V_{ud}|^2=0.94803(62)$ given in the recent review of the particle data group~\cite{2024-PDG} based on the latest critical review~\cite{2020-HT}, which, as we note here, does not account for the strong dependence on FNS effects.
The respective values of $|V_{ud}|$ are shown in Fig.~\ref{fig:CKM}.

\textit{Impact on Cabibbo unitarity---}
Since $|V_{ub}|^2\approx1.5\times10^{-5}$~\cite{2024-PDG} is too small to be relevant in the top-row CKM unitarity test, the two-flavor Cabibbo unitarity implies $|V_{ud}|=\sqrt{1-|V_{us}|^2}
={[1+|V_{us}/V_{ud}|^2]}^{-1/2}
$. This relationship is plotted in Fig.~\ref{fig:CKM}.
Currently, there are two competitive ways to determine $|V_{us}|$.
The first method is direct, using semileptonic kaon decays and a form factor calculated with lattice QCD. 
The second method uses leptonic kaon and pion decays resulting in the ratio $|V_{us}|/|V_{ud}|$.
We take $|V_{us}|=0.22330(53)$ for the first method and $|V_{us}|/|V_{ud}|=0.23108(51)$ for the second, as quoted in Ref.~\cite{2023-CKM}.
The two corresponding bands are shown in Fig.~\ref{fig:CKM}.
The updated value of $|V_{ud}|$ in this work agrees with unitarity within $1.1\,\sigma$ when considering the results of the leptonic decays and is only under slight tension of $2.3\,\sigma$ when considering the results from semileptonic kaon decay.
In both cases, the result of the analysis of this work brings us closer to unitarity by one standard deviation.

\textit{Conclusions---}In this work we revisited the extraction of the $\mathcal{F}$t value of the $^{26m}$Al$\rightarrow^{26}$Mg decay. We used the data-driven method advocated in Refs.~\cite{Seng:2022epj,Seng:2023cgl}, which employs charge radii of the members of the superallowed isotriplet as input. 
First, we parameterized the dependence of $\mathcal{F}$t on the respective nuclear charge radii of both participating nuclei, considering both the known dependence of the isospin symmetry-breaking correction and the recently identified dependence of the statistical rate function~\cite{Seng:2023cgl}. 
In doing so, we found a four-fold stronger effect than previously assumed~\cite{2020-HT, 2023-Al}.
Further, we updated the reference radii of both elements, finding shifted values compared to the literature due to the novel effect of nucleon polarization~\cite{2025-MG}.
The difference between the radii of $^{26m}$Al and $^{27}$Al was determined from the measured IS combined with a dedicated atomic many-body theory calculation of the IS factors. 
Based on the updated radii and parameterization, we revised the $\mathcal{F}t$ value for this transition and the corresponding value of $|V_{ud}|$ based solely on it, considerably reducing the CKM unitarity deficit.

\hspace{10 mm}
\begin{acknowledgments}
B.K.S. acknowledges ANRF for grant no. CRG/2023/002558 and he was supported by the Department of Space, Government of India, to carry out this work. All atomic calculations reported in the present work were computed using the ParamVikram-1000 HPC cluster of the Physical Research Laboratory (PRL), Ahmedabad, Gujarat. B.O. is grateful for the support of the Council for Higher Education Program for Hiring Outstanding Faculty Members in Quantum Science and Technology. M.~G. acknowledges support by the EU Horizon 2020 research and innovation program, STRONG-2020 project under grant agreement No 824093, and by the Deutsche Forschungsgemeinschaft (DFG) under grant agreement GO 2604/3-1. The work of C.-Y.S. is supported in part by the U.S. Department of Energy (DOE), Office of Science, Office of Nuclear Physics, under the FRIB Theory Alliance award DE-SC0013617, and by the DOE grant DE-FG02-97ER41014. We acknowledge support from the DOE Topical Collaboration ``Nuclear Theory for New Physics'', award No. DE-SC0023663. 
\end{acknowledgments} 

\bibliography{references}

\newpage
\clearpage
\appendix

\section{Atomic many-body calculations in Al-I}\label{appB}

We employ Fock-space-based single-reference relativistic coupled-cluster (FS-SR-RCC) theory to determine the atomic wave functions and energies for the ground state of Al-I, $3p ~ ^2P_{1/2}$, its fine-structure partner state $3p ~ ^2P_{3/2}$, and the excited state $4s ~ ^2S_{1/2}$, by supplementing the respective valence orbitals with the closed core $[2p^6] ~3s^2$.
In the FS-SR-RCC method, correlations are accounted for among all electrons, encompassing the core-core, core-valence, and valence-valence effects on equal footing.
Calculations are performed up to triple excitations (RCCSDT method).
For the evaluation of the IS factors, we adopt the finite-field (FF) approach.
Details of the calculation method are given in our recent review~\cite{2024-review}.

\begin{table}[hbp]
\caption{
Calculated EAs of the considered states in Al at different levels of approximation. The estimated excitation energies (EEs) are also quoted. Unless otherwise stated, all values are in cm$^{-1}$.
Our final results are compared with the experimental values (denoted Exp.).
The uncertainty for the is taken as the quadratic sum of half of the extrapolation value (+Extra) and the magnitude of missing quadruple excitations calculated in Ref.~\cite{Leonid}.
}
\begin{ruledtabular}
\begin{tabular}{l lll}
Method  & $3p~^2P_{1/2}$  & $3p~^2P_{3/2}$ & $4s^2S_{1/2}$ \\
\hline \\
DHF & 42823.87 &  42714.35 & 21311.66  \\
RMBPT(2) & 48637.77 & 48514.21 & 22760.10 \\
RCCSD & 47841.74 & 47725.64 & 22849.31 \\
RCCSDT  & 48223.08 & 48114.72 & 22914.98 \\
$+$Extra & 37.88 & 37.73 & 3.18 \\
$+$Breit & $-7.65$ & $-1.16$ & $-1.04$\\
$+$VP & $-0.26$  & $-0.25$  & 0.05 \\
$+$SE  & 4.46 & 4.18 & $-0.74$ \\
 &      &      &   \\  
Final   & 48258(20) ~\, & 48155(20) ~\, & 22916(15) ~\, \\        
Exp~\cite{1991-AlIon, 1979-LEvels} &   48278.48(3) & 48166.42(3) & 22930.72(3) \\\\
\hline \\
EEs & $^2P_{1/2}$--$^2S_{1/2}$ & $^2P_{3/2}$--$^2S_{1/2}$ & $^2P_{1/2}$--$^2P_{3/2}$\\
\hline \\
This work  & 25342(25) & 25239(25) & 102(30) \\
MCDHF(CC) \cite{2016-AlI} & 25351 & 25173 & 178 \\
MCDHF(CV) \cite{2016-AlI} & 25495 & 25376 & 119  \\
RCCSDT(Q) \cite{Leonid} & 25339(10) ~ ~ & 25230(9) ~ ~ & 109 \\
Exp~\cite{1979-LEvels} & 25347.756(1) & 25235.695(1) & 112.061(1) 
\end{tabular}
\end{ruledtabular}
\label{tab1}
\end{table}

In Table \ref{tab1}, we present the calculated electron affinities (EAs) for the ground and first two excited states of Al using the Dirac-Coulomb (DC) Hamiltonian at different levels of approximations of atomic methods.
This includes DHF method, the second-order relativistic many-body perturbation theory [RMBPT(2)], RCCSD, and RCCSDT methods.
Orbitals up to $g$-angular momentum symmetry are considered.
Contributions from the Breit interactions, lower-order vacuum polarization (VP) and self-energy (SE) effects, and `Extra' denoting contributions from the orbitals belonging to the $h$-, $i$- and $j$-angular momentum symmetries are also presented in the same table. They were estimated using the RCCSD method.
The ground-state result is compared to the experimental value and is found to agree within a few parts in $10^4$.
We also present the excitation energies (EE) for the $D1$ and $D2$ lines and the fine-structure splitting of the ground state and compare them with experiments and other calculations.
In Ref.~\cite{2016-AlI}, the MCDHF method was used with core-valence (CV) and core-core (CC) correlation frameworks, resulting in different energies.
Our results agree with those of Ref.~\cite{Leonid} and with experiment to within a few parts in $10^4$. This indicates that both CC methods, which are based on very different numerical procedures, are a good starting point to calculate IS factors.

In Table~\ref{FF:FS}, we present the FS factors calculated for the same states.
Trends in the results from the lower-order method to the higher-order method suggest that electron correlation effects play prominent roles in estimating the FS factors. Corrections from the high-lying orbitals, given as $+$Extra, and from the high-order relativistic effects are at a negligible percent level.
We also tested the effect of the assumed charge distribution on the FS factors. The results are given in Table~\ref{FS:charge} and show that there is a non-negligible effect, discussed in the main text.

\begin{table}[tbp]
\caption{
Calculated FS factors in MHz fm$^{-1}$ at different levels of approximation.
The QED contributions are roughly assessed, so that their full size is taken as uncertainty.
}
\begin{ruledtabular}
\begin{tabular}{l rrr}
Method  & $3p~^2P_{1/2}$  & $3p~^2P_{3/2}$ & $4s^2S_{1/2}$ \\
\hline \\
 DHF  & 59.22 & 59.31 & $-12.44$  \\
RMBPT(2) & 70.20 & 70.29 & $-10.71$  \\
RCCSD &  63.72 & 63.86 & $-10.90$ \\
RCCSDT & 67.53 & 67.84 & $-10.19$  \\
$+$Extra & 0.04 & 0.03 & $-0.01$\\
$+$Breit & $-0.08$ & 0.14 & $-0.10$ \\ 
$+$VP  &  $(0.07)$  & $(0.25)$  & $(0.15)$ \\
$+$SE  &  $(0.59)$  & $(0.37)$  & $(0.17)$ \\
        &      &      &   \\  
Final   & 67.5(6) & 68.0(5) & $-10.3(3)$ \\ 
\end{tabular}
\end{ruledtabular}
\label{FF:FS}
\end{table}

\begin{table*}[tbp]
\caption{
Calculated FS factors (in MHz/fm$^2$) using different charge distribution and density of electron at the origin approximations using DC Hamiltonian in the FF approach of the DHF, RMBPT(2) and RCCSD method. 
}
\begin{ruledtabular}
\begin{tabular}{l rrr rrr rrr}
Model  & \multicolumn{3}{c}{$3p~^2P_{1/2}$}  & \multicolumn{3}{c}{$3p~^2P_{3/2}$ } & \multicolumn{3}{c}{$4s^2S_{1/2}$ } \\
\cline{2-4} \cline{5-7} \cline{8-10} \\
 & DHF & RMBPT(2) & RCCSD &  DHF & RMBPT(2) & RCCSD &  DHF & RMBPT(2) & RCCSD \\
\hline \\
$\rho(0)$ & 60.38 & 71.69 & 65.16 & 59.69 & 71.04 & 64.70 & $-11.93$ & $-10.13$ & $-10.17$ \\
 Uniform  & 59.34 & 70.57 & 63.91 & 59.46 & 70.76 & 64.04 & $-12.44$ & $-10.33$ & $-10.52$\\
Gaussian & 59.40  &  70.38 & 63.89 & 58.63 & 69.65 & 63.23 & $-12.47$ & $-10.76$ & $-10.95$\\
Fermi & 59.22 & 70.20 & 63.72 & 59.31 & 70.29 & 63.86 & $-12.44$ & $-10.71$ & $-10.90$
\end{tabular}
\end{ruledtabular}
\label{FS:charge}
\end{table*}

\begin{table}[tbp]
\caption{
Calculated NMS factors (in GHz amu) with the FF approach.
The QED contributions are roughly assessed, so that their size is taken as uncertainty.
}
\begin{ruledtabular}
\begin{tabular}{l rrr}
Method  & $3p~^2P_{1/2}$  & $3p~^2P_{3/2}$ & $4s^2S_{1/2}$ \\
\hline \\
 DHF  &  703.20 & 702.28 & 350.17 \\
 RMBPT(2) & 798.79 & 797.72 & 373.96 \\
RCCSD &  785.58 & 784.58 & 375.44 \\
RCCSDT & 791.91 & 791.04 & 376.50 \\ 
$+$Extra & 0.59  &  0.58  & 0.05  \\
$+$Breit & 0.14 & 0.23 & $-0.04$\\
$+$VP  &  (1.02)  & (0.80) & (0.23) \\
$+$SE  &  (0.48)  & (0.17) & (0.04) \\
        &      &      &   \\  
Final   & 793(2)  & 792(1) &  376.5(5) \\
Scaling & 794 ~ ~ & 794 ~ ~ &  377.1 ~ ~ \\
\end{tabular}
\end{ruledtabular}
\label{FF:NMS}
\end{table}

\begin{table}[tbp]
\caption{
Calculated SMS factors (in GHz amu) with the FF approach.
The QED contributions are roughly assessed, so that their size is taken as uncertainty. It is added in quadrature with half of the extrapolated value and Breit term.
}
\begin{ruledtabular}
\begin{tabular}{l rrr}
Method  & $3p~^2P_{1/2}$  & $3p~^2P_{3/2}$ & $4s^2S_{1/2}$ \\
\hline \\
 DHF  & $-736.93$  & $-742.37$ & $-13.68$ \\
 RMBPT(2) & $-491.51$ & $-499.01$ & 36.02 \\
RCCSD &  $-589.84$ & $-597.02$  &  50.96 \\
RCCSDT  & $-597.14$ & $-599.83$  & 52.75 \\ 
$+$Extra &  $-1.09$ & $-1.10$ & $-0.03$\\
$+$Breit & 0.43 & 0.06 & $-0.02$ \\
$+$VP  & (0.08)  &(0.32) & (0.02)\\
$+$SE &  (0.64) & (1.37) & (0.64) \\
        &      &      &   \\  
Final   & $-598(2)$ & $-601(1)$ & 53(1) \\ 
\end{tabular}
\end{ruledtabular}
\label{FF:SMS}
\end{table}

In Tables~\ref{FF:NMS} and~\ref{FF:SMS}, we present the NMS and SMS factors, respectively.
The NMS factors agree well with the simple scaling law, which is only exact in the nonrelativistic framework, demonstrating that relativistic effects are small in such a light system.
The trends of the SMS factors at different level of approximations are different from the overall trend seen for the FS and NMS factors.
For both energies and IS factors, a faster convergence is seen for the $^2P_{3/2}$ and $4S$ states than the $^2P_{1/2}$ state. This indicates that the $D2$ transition IS factors are in better control than the $D1$ transition.

The results of transitions are given in Table~\ref{tab:ISF}.
To compare our results for $F$ with those of Ref.~\cite{Leonid}, we give them by calculating with a Gaussian charge distribution. The difference found is nearly $1\,\%$ for the transition of interest, much smaller than the accuracy goal for $F$ for Al-I, but much larger than the QED effects and higher-order quadrupole excitations.
As detailed in Ref.~\cite{2024-review}, the variation between the $F$ values calculated with different implementations of the RCC method is indicative of the magnitude of nonlinear field-shift effects beyond the approximation of Eq.~(\ref{eq:IS}).
For this work, it is sufficient to report the mean as a recommended value and half of the difference as uncertainty. 
Having converged on a value that includes an \textit{external} uncertainty estimation, that is, derived from independent calculations using different methods, we add the difference that arises from using a more realistic Fermi distribution, with half its value as uncertainty, and finally add the QED contribution calculated in Ref.~\cite{Leonid}.

Both the NMS and FS factors are calculated using one-body operators, so they are expected to exhibit a similar convergence pattern to that of the energies, which are reproduced with high accuracy. In accordance, remarkable agreement to within $0.1\%$ is seen between $K^\text{NMS}$ calculated in this work and in~\cite{Leonid}.
To account for external uncertainties, we report the mean value obtained in both methods with a confidence interval that spans both values.
In addition, the mean $K^\text{NMS}$ agrees with the nonrelativistic value obtained by scaling the experimental energies by the electron mass to within the validity of this approximation, which is of the order of the scaling of the fine structure energy.
Our final recommended value for $K^\text{NMS}$ is obtained by adding the QED correction of Ref~\cite{Leonid}, which is not negligible compared to the external uncertainty. Adding the external and QED correction uncertainty in quadrature results in a total uncertainty for $K^\text{NMS}$ of the two transitions at the level of $0.1\%$.

In light systems, the difficulty of calculating $K^\text{SMS}$, which is based on a two-body operator, is often the bottleneck to extracting accurate radii differences from IS measurements~\cite{2012-Co}.
Here, also, the electron correlations are strong and the numerical convergence is slow.
Nevertheless, the agreement between the atomic and molecular codes is superb, an unprecedented $0.3\%$ for the transition of interest. 
As before, we report on the mean value obtained by the two methods with a confidence interval that spans both values, and then add the QED correction of Ref~\cite{Leonid} with uncertainties added in quadrature.

\begin{table}
\caption{
Calculated isotope-shift factors using the finite-field method with the Dirac-Coulomb-Breit Hamiltonian. Values for the individual states and their dependence on the nuclear model are given in Appendix~\ref{appB}.
The uncertainty in the mean (`external') covers both values. The `internal' uncertainty in $F$ is taken as half of the nuclear model dependence.
The uncertainty in the scaling-law value is taking from scaling the fine-structure. 
Other sources of uncertainty are negligible compared to these.
}
\begin{ruledtabular}\begin{tabular}{l rr}\label{tab:ISF}

   & $^2P_{1/2}$--$^2S_{1/2}$ & $^2P_{3/2}$--$^2S_{1/2}$ \\
\hline \\
$F~$MHz/fm$^2$\\
Atomic code (TW)                & 78.0 ~ ~ & 77.7 ~ ~ \\
Molecular code (\cite{Leonid})  & 76.9 ~ ~ & 76.9 ~ ~ \\
Mean                        & 77.5(5)  & 77.3(4)\\
$\Delta$(Fermi-Gauss)       & $-0.2(1)$   & $0.6(3)$  \\
$\Delta$(QED)~\cite{Leonid} & $-0.1$ ~ ~ & $-0.1$ ~~\, \\
Recommended                 & $77.1(6)$ & $77.8(5)$ \\

\hline \\
$K^\text{NMS}$~GHz\,u\\
Atomic code (TW)                & 416.1 ~ ~ & 415.3 ~ ~ \\
Molecular code (\cite{Leonid})  & 415.6 ~ ~ & 414.9 ~ ~ \\
Mean                        & 415.9(3)  & 415.1(2)\\
Scaling                     & 417(2)\, ~ & 415(2) ~ \\
$\Delta$(QED)~\cite{Leonid} & $-1.0(3)$ & $-1.0(3)$ \\
Recommended                 & 414.9(4) & 414.1(4) \\

\hline \\
$K^\text{SMS}$~GHz\,u\\
Atomic code (TW)               & $-650.5$ ~ ~ ~ & $-653.6$ ~ ~ ~ \\
Molecular code (\cite{Leonid}) & $-654.6$ ~ ~ ~ & $-655.8$ ~ ~ ~ \\
Mean                        & $-652.6(2.0)$  & $-654.7(1.1)$ \\
$\Delta$(QED)~\cite{Leonid} & $-0.2(1)$ ~\,  & $-0.2(1)$ ~\, \\
Recommended                 & $-652.8(2.1)$ & $-654.9(1.1)$ \\

\end{tabular}\end{ruledtabular}\end{table}

\end{document}